 \documentstyle[12pt]{article} 
\begin{document} 

 \vspace*{3cm} 
 \begin{center} 
 {\LARGE \bf On The Vaidya Limit of the Tolman Model} \\[2cm] 

 {\large Charles Hellaby} \\[1mm] 
 E-mail: {\tt cwh@maths.uct.ac.za} \\[1mm] 
 {\small 
 Department of Applied Mathematics, \\ 
 University of Cape Town, \\ 
 Rondebosch, \\ 
 7700, \\ 
 South Africa} \\[7mm] 

 \vfill 
 {\it Phys. Rev. D}, {\bf 49}, 6484-8 (1994). \\
 gr-qc/9907074
 \vfill 

 {\large \bf Abstract} \\[6mm] 
 \parbox{12cm}{We show that the only Tolman models which permit a 
Vaidya limit are those having a dust distribution that is hollow 
 --- such as the 
 self-similar case.  Thus the naked 
 shell-focussing singularities found in Tolman models that are dense 
through the origin have no Vaidya equivalent.  This also casts light 
on the nature of the Vaidya metric.  We point out a hidden assumption 
in Lemos' demonstration that the Vaidya metric is a null limit of the 
Tolman metric, and in generalising his result, we find that a 
different transformation of coordinates is required.  
 } 

 \vfill 


 \vfill 

 \end{center} 

 \newpage 

 \noindent {\large \bf Introduction} 

     Recently Lemos [1992] showed that the Vaidya metric, describing 
radially directed incoherent radiation (spherically symmetric null 
dust), can be obtained from the Tolman metric, which represents a 
spherically symmetric distribution of pressureless matter (dust), by 
taking a null limit.  This surprising and intriguing insight was 
inspired by the very strong similarities, quantitative as well as 
qualitative, between the naked 
 shell-focussing singularities (discovered by Eardley and Smarr 
[1979]) that appear in the 
 self-similar forms of these metrics at the moment the crunch 
singularity forms 
 [see for example Christodoulou 1984, Newmann 1986, Ori and Piran 
1987, Rajagopal and Lake 1987, Hellaby and Lake 1988, Waugh and Lake 
1988, 1989, Grillo 1991, Lemos 1991, and extensive references in 
footnote 2 of Lake 1992]. 
 We discuss the nature of an origin of spherical coordinates in Tolman 
models, and show that a Vaidya limit cannot be extended to such a 
point.  We show that one of Lemos' assumptions can be relaxed if a 
different coordinate transformation is used. 

     The incoming Vaidya metric [Vaidya 1951, 1953, see also Lindquist 
Schwarz and Misner 1965] is 
 \begin{equation} 
  ds^2 = 2 dv dR - \left(1 - \frac{2M}{R}\right) dv^2 + R^2 d\Omega^2  
\label{eq:Vaim}  \end{equation} 
 where $d\Omega^2 = d\theta^2 + \sin^2(\theta) d\phi^2$ is the metric 
on a 
 2-sphere, the areal radius is positive, $R > 0$, and $M = M(v) > 0$ 
is an arbitrary function of the null coordinate $v$, representing the 
effective gravitational mass inside $v$.  The only 
 non-zero Einstein tensor component and the Kretschmann scalar $K = 
R^{\alpha \beta \gamma \delta} R_{\alpha \beta \gamma \delta}$ are 
 \begin{eqnarray}  G^V_{vv} & = & \frac{2}{R^2} M^*  \label{eq:Vden}  
\\ 
         K^V & = & \frac{48 M^2}{r^6}  \label{eq:VKret}  
\end{eqnarray} 
 where ${}^* = \frac{\partial}{\partial v}$, and superscripts $V$ and 
$T$ are used where necessary to distinguish quantities in the Vaidya 
and Tolman models. 

     The Tolman metric [Lema\^{\i}tre 1933, Tolman 1934] uses 
synchronous coordinates that are comoving with the dust particles, 
 \begin{equation} ds^2 = -dt^2 + \frac{R'^2}{1 + f} dr^2 + R^2 
d\Omega^2 
            \label{eq:Tolm} \end{equation} 
 where ${}' \equiv \frac{\partial}{\partial r}$, $f = f(r)$ is an 
arbitrary function of coordinate radius that determines the local 
spatial geometry, as a function of $r$ [see Hellaby and Lake 1985, 
Hellaby 1987].  The areal radius $R = R(t, r)$ is a solution of 
 \begin{equation}  \dot{R}^2 = \frac{2M}{R} + f \label{eq:Rdot2}  
\end{equation} 
 where $\dot{{}} \equiv \frac{\partial}{\partial t}$, and $M = M(r) > 
0$ is a second arbitrary function.  Comparing this equation with its 
Newtonian analogue for the kinetic plus potential energy of a radially 
moving particle of mass $m$ at a distance $x$ from the centre of a 
spherically symmetric dust cloud with density distribution $\rho_N(x)$ 
 \begin{equation} 
  \frac{m}{2} \left( \dot{x}^2 - \frac{2 M_N(x)}{x} \right) = E  
\end{equation} 
 where 
 \begin{equation}  M_N(x) = \int_0^x 4 \pi x^2 \rho_N(x) dx 
 \end{equation} 
 we obtain the interpretation that $M(r)$ is the gravitational mass 
within 
 co-moving radius $r$, and $f(r)$ is twice the energy per unit mass of 
the particles at $r$.  (The principal difference between these two 
equations is the replacement of the radial distance $x$ by the areal 
radius $R$.)  For $f > 0$ (or rather $Rf/M > 0$) the evolution of the 
areal radius for a collapsing model is hyperbolic 
 \begin{eqnarray} R & = & \frac{M}{f} (\cosh \eta - 1) 
\label{eq:hyevR} \\ 
   (\sinh \eta - \eta) & = & \frac{f^{3/2}(a - t)}{M} 
   \label{eq:hyevt} \end{eqnarray} 
 where the third arbitrary function $a = a(r)$ gives the time at which 
$R = 0$ 
 --- the big crunch.  (Parabolic and elliptic solutions exist for $f = 
0$ and $f < 0$.)  Since the pressure is zero, the dust particles 
(which stay at constant $r,\theta, \phi$) follow geodesics of the 
spacetime. It can be shown in general [Hellaby and Lake 1984, 1985] 
that for the collapsing models 
 \begin{equation} 
  R' = \left( \frac{M'}{M} - \frac{f'}{f} \right) R + \left[ a' - 
\left( 
      \frac{M'}{M} - \frac{3f'}{2f} \right) (a - t) \right] \dot{R}  
\label{eq:Rprm}  
 \end{equation} 
 The density and the Kretschmann scalar are given by 
 \begin{eqnarray} 
     8\pi \rho_T = G^T_{tt} & = & \frac{2M'}{R^2 R'} \label{eq:Tden} 
\\ 
        K^T & = & \frac{48 M^2}{R^6} - \frac{32 M M'}{R^5 R'} 
         + \frac{12 M'^2}{R^4 R'^2}  \label{eq:TKret}  \end{eqnarray} 

 \vspace{1cm}{\noindent \large \bf Lemos' Method} 

     We here outline the approach used by Lemos, although we find it 
convenient to delay taking the null limit until a slightly later stage 
in the working.  He initially makes the assumption of 
 self-similarity in both metrics, for simplicity, and later drops it.  
That assumption is not made here.  The Tolman line element 
(\ref{eq:Tolm}) may be transformed from coordinates $(t, r)$ to $(t, 
R)$ by means of 
 \begin{equation} 
  dR = \dot{R} dt + R' dr \mbox{~~~~~~~~} \rightarrow \mbox{~~~~~~~~} 
R' dr = dR - \dot{R} dt  \label{eq:Ltran1}  
 \end{equation} 
 which leads to 
 \begin{equation} 
 ds^2 =  - \left(1 - \frac{2M}{R}\right) \frac{dt^2}{(1 + f)} - 
\frac{2 \dot{R}}{(1 + f)} dt dR 
          + \frac{dR^2}{(1 + f)} + R^2 d\Omega^2 
  \label{eq:LTmetb}  \end{equation} 
 where the new $g_{tt}$ has been simplified using (\ref{eq:Rdot2}).  

     Now the limit of interest is that in which $f$ is allowed to 
diverge, while $M$ and $R$ are both required to remain finite 
 \begin{equation} 
  f \rightarrow +\infty \mbox{~~,~~~~~} 0 \leq R, M < \infty  
\label{eq:flim}  \end{equation} 
 Eq (\ref{eq:hyevR}) shows that in this limit $\cosh \eta$ must also 
diverge, so that $\cosh \eta \rightarrow \sinh \eta \rightarrow 
e^\eta/2$ and (\ref{eq:hyevR}) plus (\ref{eq:hyevt}) simplify to 
 \begin{equation}  R \rightarrow \sqrt{f} (a - t)  \label{eq:Rlim}  
\end{equation} 
 and, for finite $R$, $(a - t)$ must be vanishingly small.  Similarly 
(\ref{eq:Rdot2}) for collapsing models becomes 
 \begin{equation}  \dot{R} \rightarrow - \sqrt{f}  \label{eq:Rdlim}  
\end{equation} 
 and the derivative of (\ref{eq:Rlim}) (or alternatively substituting 
for $(a - t)$ and $\dot{R}$ from (\ref{eq:Rlim}) and (\ref{eq:Rdlim}) 
in (\ref{eq:Rprm})) gives 
 \begin{equation} 
  R' \rightarrow \frac{R f'}{2 f} + a' \sqrt{f}  \label{eq:Rplim}  
\end{equation} 

     Lemos then states that the transformation 
 \begin{equation} 
  v = \frac{t}{\sqrt{f}} + \frac{R}{f}  \label{eq:Ltran2}  
\end{equation} 
 converts (\ref{eq:LTmetb}) into the Vaidya metric (\ref{eq:Vaim}), in 
the limit $f \rightarrow\ \infty$.  Since (\ref{eq:Ltran2}) and 
(\ref{eq:Rlim}) imply that 
 \begin{equation}  v \rightarrow \frac{a}{\sqrt{f}} 
  \label{eq:Ltran2b}  \end{equation} 
 the new coordinate becomes a function of $r$ only, in the limit, so 
we can now write $M \rightarrow M(v)$. 

     We note however that a constant $f$, inherited from the 
 self-similar case, must still be assumed in order to get this result.  
If we don't make this assumption, then (\ref{eq:Ltran2}) leads to 
 \begin{eqnarray}  dv & = & \frac{dt}{\sqrt{f}} + \frac{dR}{f} 
     - \left( \frac{t}{2 f^{3/2}} + \frac{R}{f^2} \right) f' dr  \\ 
    & = & \frac{dt}{\sqrt{f}} + \frac{dR}{f} - \left( \frac{t}{2 
f^{3/2}} 
      + \frac{R}{f^2} \right) f' \frac{(dR - \dot{R} dt)}{R'}  
\end{eqnarray} 
 and, after substituting for $t$, $\dot{R}$, and $R'$ from 
 (\ref{eq:Rlim})-(\ref{eq:Rplim}), to 
 \begin{equation} 
  dv \rightarrow \left( \frac{dt}{\sqrt{f}} + \frac{dR}{f} \right) (1 
- X) 
                    \label{eq:dLtran2}  \end{equation} 
 where 
 \begin{equation} 
  X = \frac{R + a \sqrt{f}}{R + 2 f^{3/2} (a'/f')}  \end{equation} 
 so that (\ref{eq:LTmetb}) in the limit becomes 
 \begin{eqnarray}  ds^2 & \rightarrow & - \left( \frac{f}{1 + f} 
\right) 
      \left[ 2 + \frac{1}{f} \left(1 - \frac{2M}{R}\right) \right] 
dR^2 
      + \left( \frac{f}{1 + f} \right) \left( \frac{2}{1 - X} \right) 
           \left[ 1 + \frac{1}{f} \left(1 - \frac{2M}{R}\right) 
\right] dv dR  \nonumber \\ 
          && \mbox{~~~~~~~~~~} - \left( \frac{f}{1 + f} \right) 
                  \left( \frac{1}{1 - X} \right) \left(1 - 
\frac{2M}{R}\right) dv^2 
                         + R^2 d\Omega^2   \end{eqnarray} 
 The limiting behaviour of $X$ is not at all clear, as the 
relationship between $a(r)$ and $f(r)$ is arbitrary in general, and 
the limiting behaviour of $a$ is not specified. 

 \vspace{1cm}{\noindent \large \bf The Problem of the Origin and the 
Form of $f(r)$} 

     In all Tolman models describing a collapsing dust cloud which 
exhibit a naked singularity, this singularity occurs at the moment of 
collapse $t = a$, at the origin ($r = 0$ being the natural choice).  
The origin of spherical coordinates is specified by $R(t,r=0) = 0, 
\forall\; t$ and we also have $\dot{R}(t,r=0) = 0, \forall\; t$, 
which, by eq (\ref{eq:Rdot2}), requires $M(0) = 0$ as well as $f(0) = 
0$ at the origin, for example the homogeneous case (dust FLRW).  Can 
we extend Lemos' result for the null limit to cases where $f$ does not 
diverge at the origin?  Clearly the functional form of $f(r)$ must 
allow $f(0) = 0$ 
 --- for example $f = pr^2$, $p \rightarrow \infty$. 

     Consider cases with $f$ finite at $r = 0$, such as the 
 non-parabolic self-similar models.  Assuming $R,M \geq 0$, eq 
(\ref{eq:Rdot2}) shows that $\dot{R}(t, r=0) \geq f \neq 0$.  If $r = 
0$ is approached along a constant $t$ surface, with $a(r)$ finite near 
$r = 0$, (\ref{eq:hyevR}) and (\ref{eq:hyevt}) show that either (a) $M 
\rightarrow 0$ so that $\eta \rightarrow \infty$ and $R \rightarrow 
\sqrt{f} t$, or (b) $M$ remains finite so that $\eta$ and $R$ also 
remain finite.  Case (a) represents a hollow dust cloud 
 --- it can be matched at $r = 0$ onto a vacuum Tolman (Minkowski) 
spacetime with $M(r) = 0$ and a true origin at some negative $r$ value 
where $f = 0$.  Case (b) either (i) contains more dust inside $r = 0$, 
with the true origin again at $f = 0$ 
 --- i.e. $r = 0$ is not the centre of the cloud, or (ii) it contains 
the Schwarzschild vacuum inside $r = 0$, with no origin, or (iii) it 
contains a dust filled version of the 
 Schwarzschild-Kruskal-Szekeres topology [Hellaby 1987].  In (ii) and 
(iii), $f$ must pass through zero and reach $-1$ in order to form the 
throat, rising to $f \geq 0$ in the second sheet, and $M,R < \infty$ 
everywhere that $f < 0$.  Clearly particle worldlines having $f(0) > 
0$ are not at the origin, but they do collapse to zero and begin the 
formation of the singularity. 

     Furthermore, since 
 shell-focussing singularities do form in Tolman models with normal 
origins, can the detailed similarity between the naked singularities 
of the two metrics be extended to such cases, or is constant $f$ 
required? 

     Note also that the coordinate $r$ is eliminated by the first 
transformation (\ref{eq:Ltran1}) and then effectively 
 re-introduced, in the limit $v = a(r)/\sqrt{f(r)}$, via the second 
one (\ref{eq:Ltran2}).  Since the Tolman coordinate $r$ is 
 co-moving with the dust particles, and the Vaidya coordinate $v$ is 
 co-moving with the shells of radiation, one might expect $v$ to be 
the direct limit of $r$.  This is consistent with the interpretation 
of $f$ as an energy parameter which goes to infinity, meaning that the 
Tolman particle geodesics are asymptotically null.  Since a particle 
staying at the origin of spherical symmetry cannot be moving at light 
speed, this suggests that a Vaidya limit is not achieveable here.  

 \vspace{1cm}{\noindent \large \bf The Null Limit for General $f$ and 
$a$} 

     Consider approaching the origin on a constant $\eta$ surface.  
Equation (\ref{eq:hyevR}) shows that $Rf/M$ remains constant, whereas 
(\ref{eq:Rdot2}) shows that both $M/R$ and $f$ go to zero there.  Thus 
the Vaidya limit could be described by 
 \begin{equation}  \frac{Rf}{M} \rightarrow \infty \mbox{~~,~~~~~} 0 
\leq R,M < \infty 
                                          \label{eq:gVlim}  
\end{equation} 
 which doesn't necessarily require $f \rightarrow \infty$ at $r = 0$.  
The limiting forms 
 (\ref{eq:Rlim})-(\ref{eq:Rplim}) of $R$, $\dot{R}$, and $R'$ are 
unchanged by this adjustment. 

     Starting again from (\ref{eq:Tolm}), we transform from $(t, r)$ 
to $(R, r)$ as our coordinates, thus substituting for $t$ rather than 
$r$, 
 \begin{equation}  dR = \dot{R} dt + R' dr \mbox{~~~~~~~~} \rightarrow 
\mbox{~~~~~~~~} dt = (dR - R' dr)/\dot{R} 
                                          \label{eq:dttran}  
\end{equation} 
 and apply (\ref{eq:Rdot2}) to simplify the resulting $g_{rr}$ 
 \begin{equation} 
 ds^2 = - \frac{1}{\dot{R}^2} dR^2 + 2 \frac{R'}{\dot{R}^2} dR dr 
   - \left(1 - \frac{2M}{R}\right) \frac{R'^2}{(1 + f) \dot{R}^2} dr^2 
+ R^2 d\Omega^2  \label{eq:Tmetb}  
 \end{equation} 
 From (\ref{eq:Rdlim}) and (\ref{eq:Rplim}), we have the following 
limiting forms of the extra factors that don't appear in 
(\ref{eq:Vaim}) 
 \begin{equation}  \frac{1}{\dot{R}^2} \rightarrow \frac{1}{f} 
\mbox{~~,~~~~~} 
    \frac{R'}{\dot{R}^2} \rightarrow \left\{ \frac{a'}{\sqrt{f}} + 
\frac{R f'}{2 f^2} \right\}
    \mbox{~~,~~~~~} \frac{R'^2}{(1 + f) \dot{R}^2} \rightarrow \left( 
\frac{f}{1 + f} \right) 
    \left\{ \frac{a'}{\sqrt{f}} + \frac{R f'}{2 f^2} \right\}^2 
 \label{eq:limfacs}  \end{equation} 
 The limiting transformation (\ref{eq:Ltran2b}) takes care of the 
first term in the curly brackets ($a'/\sqrt{f}$), but not the second 
($Rf'/2f^2$), and without knowing both $a(r)$ and $f(r)$ 
 --- i.e. $f(a)$ --- it can't be discounted.  The second term is 
dominant if 
 \begin{equation} 
  \left( \frac{R f'}{2 f^2} \right) / \left( \frac{a'}{\sqrt{f}} 
\right) 
    = \frac{R}{2 f^{3/2}} \frac{df}{da} \rightarrow \infty  
\end{equation} 

     An example of an $f(a)$ that makes the second term dominant 
almost everywhere is 
 \begin{eqnarray} 
   f & = & a \ln(p) + sin(p^n a) \mbox{~~,~~~~~} n \mbox{~const.} 
\mbox{~~,~~~~~} p 
      \rightarrow \infty  \\ 
         \frac{df/da}{f^{3/2}} & \rightarrow & 
 \frac{f + a e^{nf/a} \cos(ae^{nf/a})}{a f^{3/2}}  \end{eqnarray} 
 but this wildly oscillating form is very unrealistic.  The conditions 
for no shell crossings [Hellaby and Lake 1985] for collapsing 
hyperbolic Tolman models require $f' > 0$ and $a' > 0$ wherever $M' > 
0$, i.e. $df/da > 0$, so adding a linear term to remove negative 
gradients gives a vanishing second term 
 \begin{eqnarray} 
  f & = & 2 p^n a + sin(p^n a) \mbox{~~,~~~~~} n \mbox{~const.} 
\mbox{~~,~~~~~} p 
    \rightarrow \infty  \\ 
   \frac{df/da}{f^{3/2}} & \rightarrow & \frac{2 + \cos(f/2)}{2 a 
\sqrt{f}}  
 \end{eqnarray} 
 The most rapid uniform divergence of $df/da$ relative to $f$ we have 
been able to find for $df/da > 0$ still leaves $(df/da)/f^{3/2}$ 
vanishing.  It is expressed in terms of computer notation $\hat{\;\;}$ 
for raising to the power, 
 \begin{eqnarray}  f & = & a p\hat{\;\;}(p\hat{\;\;}(p\hat{\;\;} ... 
       (p\hat{\;\;}a))) \mbox{~~,~~~~~} p \rightarrow \infty  \\ 
  \frac{df/da}{f^{3/2}} & \rightarrow & 
   \frac{\ln(f) \ln\ln(f) \ln\ln\ln(f) ... }{a \sqrt{f}}  
\end{eqnarray} 
 However, at a single point (or a finite number of discrete points) 
the divergence behaviour can always be made arbitrarily rapid, e.g. 
 \begin{eqnarray} 
  f & = & p a + a^{p^n} \mbox{~~,~~~~~} n \mbox{~const.} 
\mbox{~~,~~~~~} p \rightarrow \infty  \\ 
  \frac{df/da}{f^{3/2}} & \rightarrow & \frac{1}{a \sqrt{f}} 
\mbox{~~,~~~~~} 0 < a < 1  \\ 
             & \rightarrow & f^{n-3/2} \mbox{~~,~~~~~} a = 1  \\ 
   & \rightarrow & \frac{\ln(f)}{\ln(a) \sqrt{f}} \mbox{~~,~~~~~} a > 
1  \end{eqnarray} 

 Consequently, we now introduce the following transformation, 
 \begin{eqnarray}  v & = & \int_0^r \frac{a'}{\sqrt{f}} dr - 
\frac{R}{2f}  
                                 \label{eq:vtran}  \\ 
         dv & = & \left( \frac{a'}{\sqrt{f}} + \frac{R f'}{2 f^2} 
\right) dr 
             - \frac{dR}{2 f}  \label{eq:dvtran}  \end{eqnarray} 
 which incorporates both terms in the brackets of (\ref{eq:limfacs}), 
and which converts (\ref{eq:Tmetb}) to 
 \begin{eqnarray} ds^2 & \rightarrow & - \frac{1}{4 f (1 + f)} \left(1 
- \frac{2M}{R}\right) dR^2 
           + \left[2 - \frac{1}{(1 + f)} \left(1 - \frac{2M}{R}\right) 
\right] dv dR  \nonumber \\ 
          && \mbox{~~~~~~~~~~} - \left( \frac{f}{1 + f} \right) 
\left(1 - \frac{2M}{R}\right) dv^2 
                         + R^2 d\Omega^2  \label{eq:Tmetc}  
\end{eqnarray} 
 It is already clear from (\ref{eq:limfacs}) as well as this equation 
that $f \rightarrow \infty$ is indeed required everywhere to obtain 
the Vaidya metric as the limit.  
 \begin{equation} ds^2 \rightarrow 2 dv dR - \left(1 - 
\frac{2M}{R}\right) dv^2 + R^2 d\Omega^2 
 \end{equation} 
 (The alternative transformation 
 \begin{equation}  v = \int_0^r \frac{a'}{\sqrt{1 + f}} dr 
      - \sqrt{\frac{1 + f}{f}}R  \mbox{~~,~~~~~}  
         dv = \sqrt{\frac{f}{1 + f}} \left( \frac{a'}{\sqrt{f}} 
      + \frac{R f'}{2 f^2} \right) dr - \sqrt{\frac{1 + f}{f}} dR 
   \end{equation} 
 does not succeed in removing the factor of $f/(1 + f)$, and also 
leads to the wrong limit.)  In the limit (\ref{eq:flim}) then, it is 
evident from (\ref{eq:vtran}) that $v$ becomes a function of $r$ only, 
so that $M \rightarrow M(v)$ holds once again.  No assumptions about 
the functional form of $f$ or the limiting behaviour of $a$ were made 
to obtain the Vaidya metric as the null limit, and we find that the 
second term of (\ref{eq:vtran}) becomes negligible, even if the second 
term in the brackets of (\ref{eq:dvtran}) doesn't.  The new 
transformation (\ref{eq:vtran}) can also be 
 re-written in the limit as 
 \begin{equation}  dv \rightarrow \frac{dt}{\sqrt{f}} + \frac{dR}{2f} 
   \label{eq:dvdtdR}   \end{equation} 
 in order to recover (\ref{eq:Vaim}) from (\ref{eq:LTmetb}).  
Equations (\ref{eq:dvdtdR}) and (\ref{eq:vtran}) are the revised 
versions of (\ref{eq:dLtran2}) and (\ref{eq:Ltran2b}). 

     The overall transformation from Tolman to asymptotically Vaidya 
coordinates then is 
 \begin{eqnarray}  v & = & \int_0^r \frac{a'(r)}{\sqrt{f(r)}} dr 
              - \frac{R(t,r)}{2 f(r)}  \label{eq:gVtranv}  \\ 
         R & = & R(t,r)  \label{eq:gVtranR}  \end{eqnarray} 
 where $R(t,r)$ is given by 
 (\ref{eq:hyevR})-(\ref{eq:hyevt}).  Using the following limiting 
values of two of the partial derivatives of the inverse transformation 
 \begin{equation} \left. \frac{\partial r}{\partial v} \right|_R = 
\frac{f}{R'} 
 \mbox{~~,~~~~~} \left. \frac{\partial t}{\partial v} \right|_R = 
\sqrt{f} \end{equation} 
 the Kretschmann scalar and the density may be converted to their 
Vaidya forms.  Thus 
 \begin{equation}  M^* = \left. \frac{\partial M}{\partial v} 
\right|_R = 
  \left. \frac{\partial M}{\partial r} \right|_t \left. \frac{\partial 
r}{\partial v} \right|_R = M' \frac{f}{R'} 
        \mbox{~~~~i.e.~~~~}    \frac{M'}{R'} = \frac{1}{f} M^* 
  \end{equation} 
 ensures that the last two terms on the right of (\ref{eq:TKret}) 
vanish, giving (\ref{eq:VKret}) in the limit.  For the `density', 
(\ref{eq:Vden}) is obtained from (\ref{eq:Tden}) in the limit by 
writing 
 \begin{equation} 
  G^V_{vv} = \left( \left. \frac{\partial t}{\partial v} \right|_R 
\right)^2 G^T_{tt} 
    = f \frac{2 M'}{R^2 R'} = \frac{2}{R^2} M^*  \end{equation} The 
strengths of singularities are variously defined by [e.g. Tipler 
Clarke and Ellis 1980, Clarke and Kr\'{o}lak 1986] 
 \begin{equation} 
  \Psi_G = lim_{\lambda \rightarrow 0} \lambda^2 G_{\alpha \beta} 
k^\alpha k^\beta 
  \mbox{~~~~~~or~~~~~~} 
  \Psi_R = lim_{\lambda \rightarrow 0} \lambda^2 R_{\alpha \beta} 
k^\alpha k^\beta 
 \end{equation} 
 where $k^\alpha$ is the tangent vector to a null geodesic with 
parameter $\lambda$ that hits the singularity at $\lambda = 0$.  From 
the above, and since $\Psi$ is a scalar, it is clear that the 
strengths of the Vaidya singularity, as measured along radial 
geodesics are given by the limits of the corresponding Tolman 
expressions. 

 \vspace{1cm}{\noindent \large \bf Conclusions} 

     Lemos originally demonstrated that the Vaidya model is a null 
limit of the Tolman model, by taking the limit $f \rightarrow \infty$ 
and assuming $f = $ constant in this limit.  His transformation was 
completely valid for models with constant $f$.  However Tolman 
 shell-focussing singularities also occur in models with matter at the 
origin.  The existence of a normal origin of spherical coordinates at 
$r = 0$, $(a - t) > 0$ in the Tolman model requires $f(r=0) = 0$, and 
we have found this cannot be made consistent with a null limit.  The 
Vaidya limit does indeed require $f \rightarrow \infty$, so it cannot 
be extended to a spherical origin, where $f(0) = 0$, or a 
 Schwarzschild-Kruskal-Szekeres type topology, which requires $f = -1$ 
in the throat. 

     Thus we conclude that every Vaidya model is the limit of a hollow 
Tolman model, acquiring its arbitrary $M(v)$ from a combination of 
$M(r)$ and $a(r)$, and must itself be hollow.  If $M(r=0) = 0$, 
$M(v=0) = 0$, then $r = 0$, $v = 0$ is a collapsing shell of finite 
size surrounding Minkowski vacuum, and the limiting Vaidya model can 
form a 
 shell-focussing.  If $M(r=0) > 0$, $M(v=0) > 0$ then it surrounds 
Schwarzschild vacuum, and no 
 shell-focussing forms.  In this latter case, the shells of incoming 
radiation (having $f$ divergent) cannot pass through the throat (where 
$f = -1$) and must hit the future singularity first.  A dust filled 
interior is not possible in the limit, since a coordinate line cannot 
be 
 co-moving with both a dust particle and a light ray, but it may be 
possible to have an intervening vacuum region.  Since $t = a$ on the 
singularity, and $R$ is only finite on a collapsing shell of radiation 
where $(a - t)$ is infinitesimal, the radiation is all at infinite $R$ 
for any finite value of $(a - t)$. 

     If we assume that $M^*$ is finite, then it is apparent from 
(\ref{eq:vtran}) that a finite change in $M$ and $v$ requires an 
infinite change in $a$.  It is interesting to note that a collapsing, 
unbound (i.e. hyperbolic) dust cloud of finite total mass may also 
have $f,a \rightarrow \infty$ and $M$ finite in the asymptotic 
regions.  At finite $(a - t)$, $R$ is infinite, from 
 (\ref{eq:hyevR})-(\ref{eq:hyevt}), but as these particles collapse 
towards the crunch, $R$ becomes finite when $(a - t)$ is 
infinitesimal, and the Vaidya limit is achieved.  In terms of Tolman 
time, this is infinitely long after the initial formation of the 
singularity, but only a finite retarded time in Vaidya coordinates. 

     The new coordinate transformation 
 (\ref{eq:gVtranv})-(\ref{eq:gVtranR}) 
 --- or (\ref{eq:dttran}) and (\ref{eq:dvtran}) 
 --- makes no assumptions about the 3 arbitrary Tolman functions $f$, 
$M$, and $a$, and in particular the relationship between $f$ and $a$, 
beyond those normally made for a general Tolman model, and the limit 
$f \rightarrow \infty$.  Several important physical quantities 
 --- the Einstein tensor, the Kretschmann scalar, the null geodesics, 
and the strenghts of the singularities 
 --- all have the correct limit.  This generalises Lemos' unification 
of the two metrics and their 
 shell-focussing singularities. 

 \newpage 
 \noindent {\large \bf References} ~\\[5mm] 

 \hspace*{12mm}\parbox{15cm}{\small 
 \setlength{\parindent}{-2cm} 
 \setlength{\parskip}{2ex} 

 Christodoulou D 1984, {\it Comm. Math. Phys.} {\bf 93}, 171-95. 

 Clarke C J S and Kr\'{o}lak K 1986, {\it J. Geom. Phys.} {\bf 2}, 
127. 

 Eardley D M and Smarr L 1979, {Phys. Rev. D} {\bf 19}, 2239-59. 

 Grillo 1991, {\it Class. Q. Grav.} {\bf 8}, 739. 

 Hellaby C W 1987, {\it Class. Q. Grav.} {\bf 4}, 635-50.

 Hellaby C W and Lake K 1984, {\it Astrophys. J.} {\bf 282}, 1-10, 
with corrections in {\it ibid} {\bf 294}, 702. 

 Hellaby C W and Lake K 1985, {\it Astrophys. J.} {\bf 290}, 381-7, 
with corrections in {\it ibid} {\bf 300}, 461. 

 Hellaby C W and Lake K 1988, {\it Preprint} ``The Singularity of 
Eardley Smarr and Christodoulou", UCT Applied Maths preprint 88/7. 

 Lake 1992, {\it Phys. Rev. Lett.} {\bf 68}, 3129-32. 

 Lema\^{\i}tre G 1933, {\it Ann. Soc. Sci. Bruxelles} {\bf A53}, 51. 

 Lemos 1991, {\it Phys. Lett. A} {\bf 158}, 279-81. 

 Lemos 1992, {\it Phys. Rev. Lett.} {\bf 68}, 1447-50. 

 Lindquist R W, Schwarz R A, and Misner C W 1965, {\it Phys. Rev.} 
{\bf 137B}, 1364. 

 Newmann R P A C 1986, {\it Class. Q. Grav.} {\bf 3}, 527. 

 Ori and Piran 1987, {\it Phys. Rev. Lett.} {\bf 59}, 2137. 

 Rajagopal K and Lake K 1987, {\it Phys Rev. D} {\bf 35}, 1531. 

 Tipler F J, Clarke C J S and Ellis G F R 1980, in {\it General 
Relativity and Gravitation}, ed. A Held (Plenum, New York). 

 Tolman R C 1934, {\it Proc. Nat. Acad. Sci.} {\bf 20}, 169. 

 Vaidya P C 1951, {\it Proc. Indian Acad. Sci.} {\bf A33}, 264. 

 Vaidya P C 1953, {\it Nature} {\bf 171}, 260. 

 Wald R M 1984, {\it General Relativity} (University of Chicago 
Press), p~153. 

 Waugh and Lake 1988, {\it Phys. Rev. D} {\bf 38}, 1315. 

 Waugh and Lake 1989, {\it Phys. Rev. D} {\bf 40}, 2137. 

 } 

 \end{document}